# Dynamic self-assembly of active particle systems controlled by light fields[*]

Si-hang Guo , Guang-yu Yang , Guo-qing Meng , Ying-ying Wang , Jun-xing PAN[*] , Jin-jun ZHANG[*]

School of Physics and Information Engineering, Shanxi Normal University, Taiyuan 030032, China

Active particle systems are a class of non-equilibrium systems composed of self-propelled Brownian particles; through interactions between particles within the system, a variety of intriguing collective behaviors can emerge. Based on Brownian dynamics simulations, this study investigates the formation and transition mechanisms of ordered structures in active particle systems regulated by light fields. The study reveals that, under light field regulation, active particles undergo large-scale phase separation behavior, forming specific ordered structures and enabling the dynamic transition between multiple ordered structures. This study systematically explores the effects of light fields on this dynamic phase transition and the corresponding regulatory mechanisms. The research findings provide important references for the precise regulation of collective structures in active systems and the fabrication of micro-nano intelligent devices.

## 1. Introduction

In nature, active matter exist widely, ranging from nano-scale bacterial colonies to large biological communities[1,2]. Such systems are driven by external energy inputs, enabling autonomous motion and exhibiting pronounced non-equilibrium dynamical behaviors [3–5]. The study of the non-equilibrium collective behavior of active matter is crucial for advancing our understanding of motion phenomena at the microscopic

---

scale and the collective dynamics of various natural communities. Moreover, this research contributes to the enrichment and expansion of the theoretical framework of non-equilibrium dynamics.

Due to the complexity of active matter, laboratories substitute active matter with the fabrication of corresponding artificial active particles[6–10], which facilitates the investigation of the collective behavior of active matter. Environmental energy acts as a power source for these particles, which they convert into self-propelled motion. Currently, extensive research has been conducted on the collective motion of active particles and their non-equilibrium dynamical behaviors under external fields, spanning theory, simulations, and experiments. Stenhammar et al. explored phase separation driven by active components and self-assembly in active-passive particle mixtures[11]. Dolai et al. investigated phase separation in binary mixtures of active and passive particles[12]. They found that the motion of small active particles induces an effective attraction between large passive particles, and this attraction drives the phase separation of large-sized particles. Gao Y W et al. conducted simulation studies on the cross-region motion of active polymers under periodic external field modulation[13]. They found that the properties of the external field play a key role in regulating the behavior of active polymers. Fernandez-Rodriguez et al. [14] achieved independent and precise control over the rotational diffusivity ($D$r) of active Brownian particles via a tunable magnetic field and a discrete-time feedback loop. This control tuned the rotational diffusivity across distinct regions, enabling them to investigate the locomotion behavior and spatial distribution characteristics of active particles in complex environments.

As an ideal power source for active systems, light fields allow precise control over their intensity and range, along with the combined attributes of remote controllability and high spatiotemporal precision. Regulating light fields enables modulation of the motile behavior of target monomers[15,16]. Zhang et al. investigated the aggregation and self-assembly of light-controlled colloidal particles (CPs)[17]. B Bäuerle et al.[18] employed photoactivated active particle suspensions, alongside particle detection algorithms and a scanning laser system, to investigate how variations in interparticle responses influence their self-assembly. The research group of Liu Liyu and their collaborators investigated the locomotion behavior of photosensitive active matter composed of robot swarms in planar dynamic resource environments—represented by large-scale light-emitting diode arrays[19]—as well as the complex interactions among the robotic swarm-based active matter and between individual robots and the environment under external field driving, which give rise to ordered hysteresis[20].

Research on the collective motion of active systems under light field actuation is still in its infancy, with many open questions remaining to be addressed—particularly regarding the large-scale dynamic self-assembly behaviors exhibited by active matter [21–30]. Thus, in this work, we employ Brownian dynamics simulations to investigate the dynamic self-organization of active particles under light field actuation. By analyzing the influence of various parameters on the system's dynamic self-organization, we establish and refine the dependence of active particle phase separation on these parameters.

## 2. Model and method

In this work, based on Brownian dynamics simulations [31–40], all particles in the simulated system adopt the active Brownian particle (ABP) model. Each active particle acts as a self-propelled sphere, with its self-propulsion force derived from the light field. Previous studies have shown that active particles can absorb light energy to achieve self-propulsion, and the average velocity of their motion increases linearly with increasing light intensity [41–44]. Fig. 1(a) further illustrates the relationship between particle velocity ($v$) and self-propulsion force ($F$) in the system studied herein. Preliminary calculations show a linear proportional relationship between the two. By analyzing the positive correlations among velocity, light intensity, and self-propulsion force, we verify the regulatory role of light intensity in particle activity. To investigate the collective motion of active particles under a periodic striped light field, we constructed a $100\sigma \times 100\sigma$ box in a two-dimensional (2D) plane. The active particles in the box had an areal density of 0.5, and a periodic striped light field with equal spacing and equal width was applied simultaneously along the x- and y-axes. This configuration resulted in alternating illuminated and shadowed regions—each with a width of $20\sigma$—along both the x- and y-axes. Owing to the superposition effect of the transverse and longitudinal light fields, a checkerboard pattern—shown in Fig. 1(b)[14]—is formed across the entire box. In the overlapping regions of transverse and longitudinal shadowed areas ((dark gray regions in Fig. 1(b)), no light illumination is present; thus, the self-propulsion force ($F$) of particles is zero, and the particles behave as ordinary Brownian particles. In all other regions, particles are subjected to self-propulsion forces induced by light illumination. Moreover, the stronger the light intensity, the larger the particle self-propulsion force; consequently, the self-propulsion force of particles is maximized in the overlapping regions of transverse and longitudinal illuminated areas (light yellow regions in Fig. 1(b)).

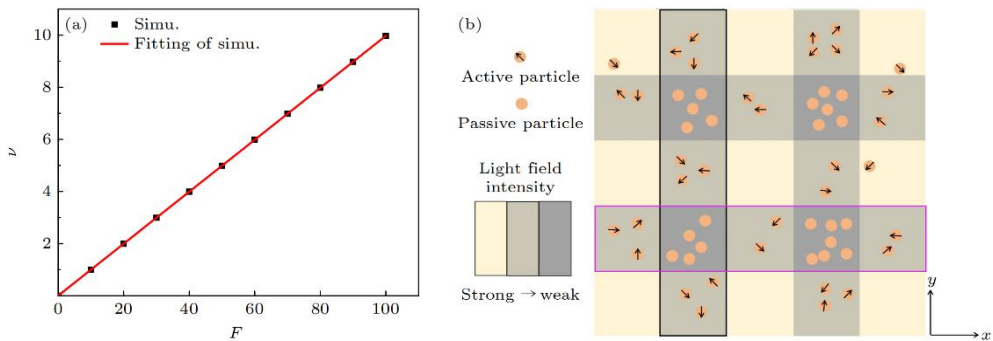

**Figure 1.** (a) Relationship between the velocity $v$ of active particles and the self-propulsive force $F$ in a uniform light field; (b) model schematic of active particles driven by the periodic

striped light field. The region marked by the black box is Zone I, representing a longitudinal shaded region, while the region marked by the purple box is Zone II, representing a transverse shaded region.

A purely repulsive Weeks-Chandler-Andersen (WCA) potential is employed to describe the interactions between all particles:

$$U_{WCA} = \begin{cases} 4\varepsilon\left[\left(\frac{\sigma}{r}\right)^{12} - \left(\frac{\sigma}{r}\right)^{6}\right] + \varepsilon, r < r_c \\ 0, r > r_c \end{cases} \quad (1)$$

Where $r_c = 2^{1/6}\sigma$ denotes the cutoff radius of the particles; $\sigma$ is the diameter of the particle; $\varepsilon$ represents the interaction strength between particles. When $r > r_c$, the interaction potential between particles is 0. The motion of each particle follows the Langevin equation of motion:

$$m\ddot{r}_i = -\frac{\partial U_i}{\partial r_i} - \zeta \dot{r}_i + F(r)\hat{u}_i(t) + \sqrt{2\zeta k_B T}\eta_i(t), \quad (2)$$

$$m\ddot{r}_i = -\frac{\partial U_i}{\partial r_i} - \zeta \dot{r}_i + F(r)\hat{u}_i(t) + \sqrt{2\zeta k_B T}\eta_i(t). \quad (3)$$

$$\dot{\theta}_i = \sqrt{2D_r}\xi_i(t). \quad (4)$$

Equation (2) describes the motion of active Brownian particles in the translational degree of freedom within illuminated regions, where $r_i$ denotes the position of the $i$-th colloidal particle; $U_i$ represents the interaction potential energy; $\zeta$ is the translational friction coefficient; $k_B T$ stands for the scaled temperature; and $\hat{u}_i = (\cos\theta, \sin\theta)$ represents the unit vector in the direction of the self-propulsion force $F$. Equation (3) represents the motion equation of Brownian particles in shadowed regions. Equation (4) represents the rotational diffusion equation, where $D_r$ is the rotational diffusion coefficient with $D_r = 3D_0/\sigma^2$ ($D_0$ is the translational diffusion coefficient). $\eta(t)$ and $\xi(t)$ represent Gaussian white noise.

In this work, the simulation was implemented using LAMMPS[45–51], with reduced units adopted: $m=1$, $\varepsilon=1$, $\sigma=1$, and $k_BT=1$; the time unit is $\tau = \sqrt{m\sigma^2/k_BT}$. In addition, the friction coefficient $\zeta=10$ is used in this paper to ensure that all particles are in an effective overdamped environment. Periodic boundary conditions were employed in the simulation.

## 3. Results and Discussion

### 3.1 Phase diagram

This work first investigates the influence of light intensity on the aggregate structure of the system under different rotational diffusion coefficients. The self-propulsion force of the transverse light field is fixed at $F_2=10$, while the self-propulsion force of the longitudinal light field $F_1$ is varied. Both the width of the illuminated region $W$ and the spacing between adjacent illuminated regions $S$ are set to $20\sigma$. The variation of the system's aggregate structure with the ratio $P = F_1/F_2$ (where $P$ is the ratio of the self-propulsion force of the longitudinal light field to that of the transverse light field) and the rotational diffusion coefficient $D_r$ is shown in Fig. 2. Fig. 2(a) presents the phase diagram illustrating the variation of the system's aggregate structure; Fig. 2(b)-(e) show the density distribution maps of representative stable structures; and Fig. 2(f) displays the time evolution of the particle number density in the longitudinal shadowed regions for different structures under various conditions.

As observed in Fig. 2(a), when the rotational diffusion coefficient is small (i.e., $D_r \leq 0.005$), the system gradually transitions from an unstable cluster structure to an oscillatory structure as the ratio $P$ increases. A small rotational diffusion coefficient prevents particles from rapidly adjusting their direction of motion to adapt to light field variations. When particles enter shadowed regions under light-driven propulsion, they collide with particles already aggregated in these shadowed regions; this collision forces the aggregated particles to be expelled from the shadowed regions. The expelled particles, driven by the light field, then move toward the shadowed regions again. This cycle results in particles frequently entering and exiting the shadowed regions, precluding the formation of stable aggregates (see Supplementary Videos SV1 and SV2 (online)). Meanwhile, when the ratio $P$ is large, the longitudinal light intensity is high, which increases the velocity of particles moving within illuminated regions[42] and leads to more frequent and intense collisions. When a large number of active particles collide with particles in the shadowed regions, the particle number density in the shadowed regions decreases sharply. Subsequently, the particles knocked into the illuminated regions rapidly return to the shadowed regions under the driving force of the strong light field, causing the density to rise again. This process forms a unique oscillatory structure, as shown by the V-curve in Fig. 2(f).

This reflects the combined effect of a strong longitudinal light field and a small rotational diffusion coefficient on particle motion. In this oscillatory structure, the particle number density in the longitudinal shadowed regions exhibits quasi-periodic fluctuations over time, with large density fluctuations. In contrast, in the unstable cluster structure, the variation in particle number density is relatively random, with small fluctuations and no obvious periodicity, as shown by the VI-curve in [Fig. 2(f)](). A comparison of the V-curve and VI-curve in [Fig. 2(f)]() reveals that the average density of the oscillatory structure is consistently higher than that of the unstable cluster structure. This is because under the action of a strong light field, particles are less likely to remain in illuminated regions—leading to a significant reduction in the number of particles in illuminated regions and an increase in the number of particles in shadowed regions.

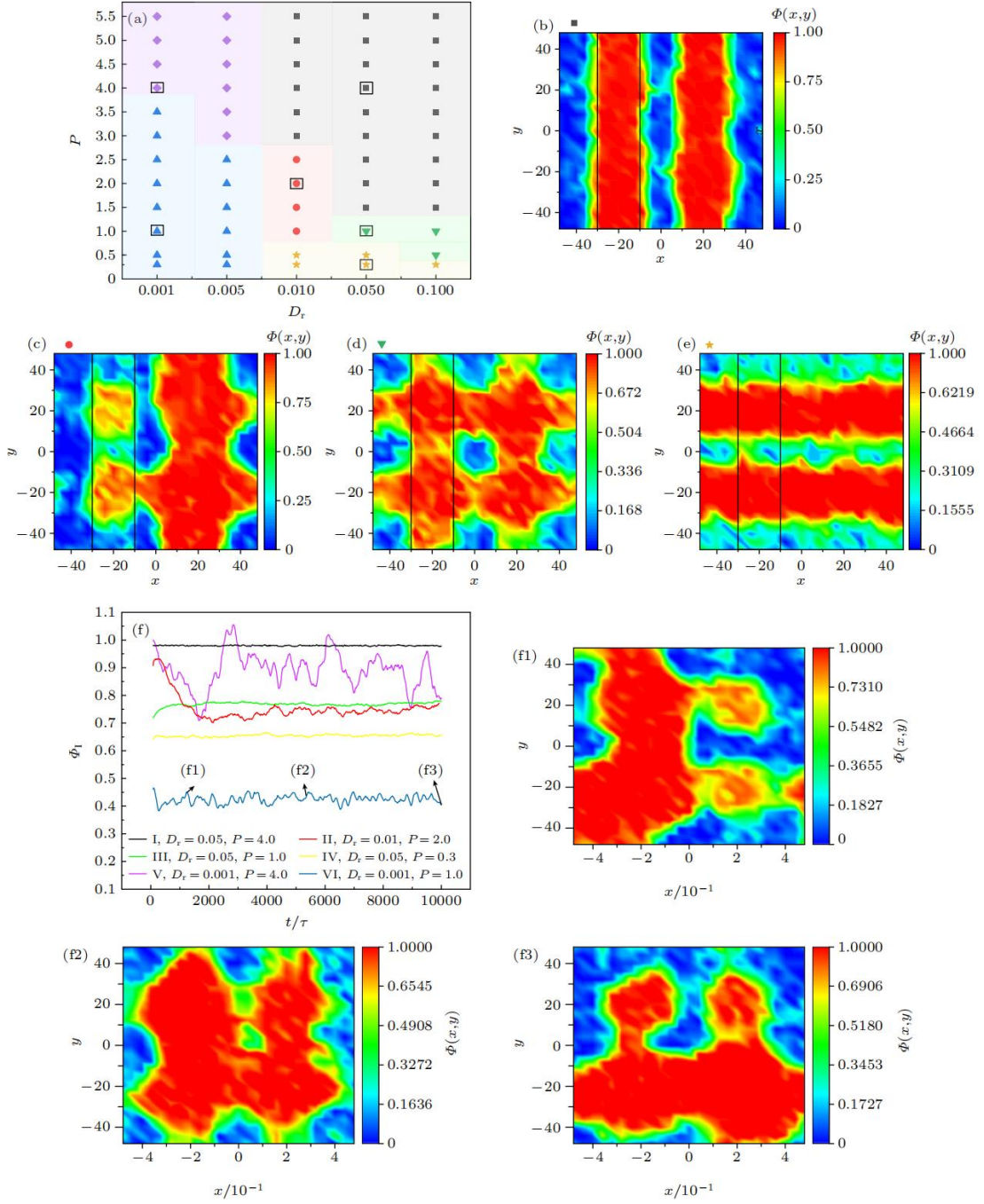

**Figure 2.** (a) Phase diagram of the system's aggregation structures. ◆ represents an oscillatory structure; ▲ represents an unstable cluster structure; ■ represents a longitudinal stripe structure; ● represents a mixed striped block-like hybrid structure; ▼ represents a tic-tac-toe structure; ★ represents a transverse stripe structure. (b), (c), (d), and (e) The density distribution of representative structures, with the black boxes indicating the longitudinal shading region I. (f) Temporal evolution of particle number density in region I for different structures.

At high rotational diffusion coefficients (i.e., $D_r \geq 0.01$), the system as a whole exhibits a more stable trend (see Supplementary Videos SV3–SV6 (online) for further details). The fundamental reason for this phenomenon lies in the fact that at large rotational diffusion coefficients, the direction of particle motion changes more rapidly. This enables particles to quickly adjust their direction to adapt to light field variations when moving between shadowed and illuminated regions, thereby reducing frequent and intense collisions and rendering the system more ordered and stable—as shown by the I–IV curves in Fig. 2(f). For these four stable structures, the variation in particle number density within the shadowed regions tends to be stable, indicating that the system has reached a steady state.

Fig. 3 shows that as the ratio $P$ increases, the particle number densities in the longitudinal shadowed regions and transverse shadowed regions undergo a reversal. This corresponds to the transition of the system's structure from transverse stripes to longitudinal stripes in the Fig. 2(a) when the rotational diffusion coefficient is large. In other words, the change in the system's steady state reflects a gradual transition of the particle distribution from being transversely dominated to longitudinally dominated. Additionally, the transition between transverse and longitudinal aggregated states exhibits distinct transition processes. At $D_r = 0.01$, the transition state of the system is a striped block-like hybrid structure; at $D_r=0.05$, however, this transition state transforms into a tic-tac-toe structure. As observed in Fig. 2(f), the stability of the striped block-like hybrid structure lies between that of stable structures and unstable structures. Thus, this structure can be interpreted not only as a transition state during the transformation from transverse stripes to longitudinal stripes, but also as a transitional structure in the process of the system's stability change. Furthermore, the striped block-like hybrid structure exhibits another equi-probable form (see Supplementary Video SV7 (online) for further details).

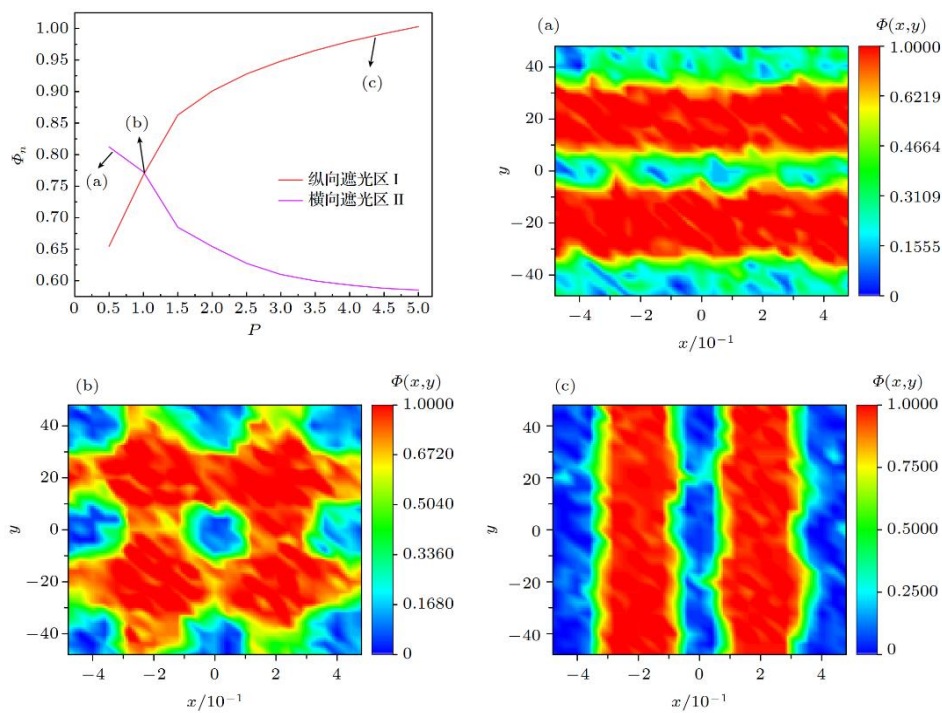

**Figure 3.** Relationship between particle number density ($\phi_n$) and the ratio $P$ of longitudinal to transverse light-induced self-propulsion forces in the longitudinal shading region I and the transverse shaded region II at $D_r = 0.05$.

Simulation results show that particles aggregate in shadowed regions and weak illuminated regions, which is consistent with the principle of motility-induced phase separation (MIPS). Taking the longitudinal stripe structure as an example, Fig. 4(a) presents the distribution profiles of the particle number density and average velocity along the x-axis in the transverse shadowed region (II). The results show that due to the weakest light field in shadowed regions, the average velocity of particles decreases significantly. This reduction in velocity causes particles to aggregate in these regions; in turn, the accumulated particles further slow down their own motion, forming a positive feedback loop. This ultimately leads to phase separation between high-density and low-density regions of the particle swarm. When the ratio is $P > 1$, the longitudinal light intensity is higher than the transverse light intensity. Particles are mainly dominated by the longitudinal light field: their velocity in the longitudinal illuminated regions is high, which inhibits aggregation in these longitudinal illuminated regions while promoting aggregation in the longitudinal shadowed regions—ultimately forming a longitudinal stripe structure. Similarly, when the ratio is $p < 1$, the transverse light field dominates particle motion, causing particles to tend to aggregate in transverse shadowed regions and form an ordered arrangement of transverse stripes. When the light intensities of the two (longitudinal and transverse light fields) are comparable with no dominant light field, particles distribute in both transverse and longitudinal shadowed regions, and the system exhibits a tic-tac-toe structure. This process reflects the regulatory effect of changes in the relative light intensity between the transverse and longitudinal directions on particle activity and aggregation behavior.

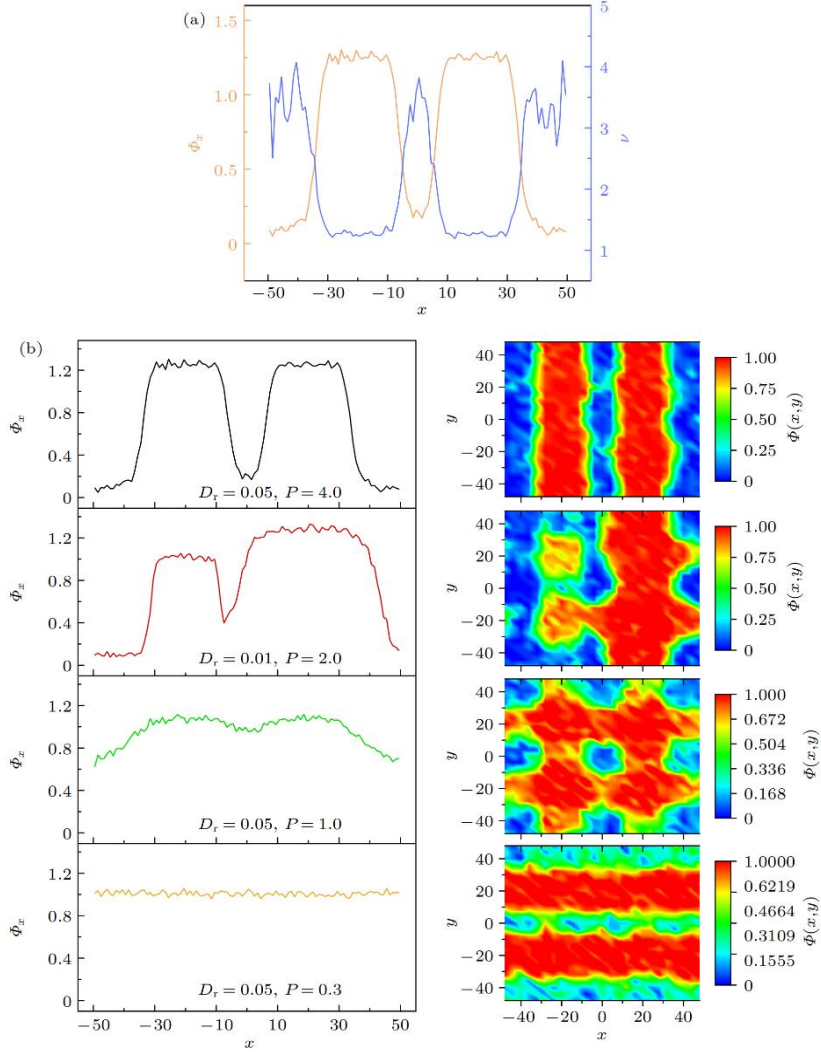

**Figure 4.** (a) Distributions of particle number density and velocity along the *x*-axis within the transverse shaded region (II) of the longitudinal stripe structure; (b) distributions of particle number density along the *x*-axis in the transverse shaded region (II) for four steady states.

Fig. 4(b) presents the spatial distribution of particle number density for the four steady-state structures. The specific manifestation of phase separation in each structure can be clearly observed: the transverse stripe structure exhibits a continuous high-density distribution, whereas the longitudinal stripe structure, striped block-like hybrid structure, and tic-tac-toe structure show a bimodal distribution—due to the enhanced particle aggregation in the overlapping regions of transverse and longitudinal shadowed regions. Among these structures, the particle number density distribution curves of the transverse stripe structure and tic-tac-toe structure exhibit minimal fluctuations. Furthermore, the bimodal distribution of the striped block-like hybrid structure differs from that of the longitudinal stripe structure, showing asymmetry: one peak has a lower intensity and narrower width, while the other has a higher intensity and wider width.

### 3.2 Dynamics

The transition of the system's aggregated structures can be analyzed from a dynamic perspective. For this reason, we calculated the mean squared displacement (MSD) of the particle centroids under different conditions, defined as $MSD = \langle |\Delta r_i(t)|^2 \rangle$ as well as the scaling exponent $\alpha(t) = \frac{d \log(MSD)}{d \log(t)}$. The log-log plot of the evolution of MSD with time for different structures and the corresponding scaling exponent $\alpha$ with time are shown in Fig. 5.

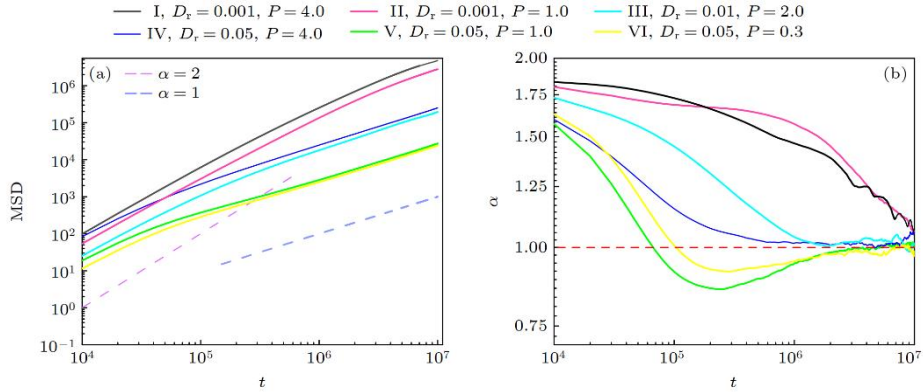

**Figure 5.** (a) Log-log plot of the mean squared displacement (MSD) of the particle center of mass as a function of time, where the dotted lines show slopes of 2.0 or 1.0; (b) scaling exponent $\alpha$ as a function of time.

In the unstable cluster structure and oscillating structure— as shown by the I and II curves in Fig. 5(b) — the scaling exponent $\alpha$ remains greater than 1 throughout the entire simulation process, exhibiting a distinct superdiffusive behavior. This indicates that particles maintain high motile activity throughout the entire evolution process. For the other four steady states— as shown by the III–VI curves in Fig. 5(b)— particles exhibit superdiffusive behavior in the early stage, which indicates a high diffusion rate of particles during this stage. As time progresses, the scaling exponent $\alpha$ of these four structures gradually approaches 1, eventually exhibiting normal diffusive behavior. This indicates that the system reaches dynamic equilibrium over longer timescales, and the particle motion gradually transitions to Brownian motion, ultimately aggregating in shadowed regions. Among these four steady structures, the scaling exponent $\alpha$ of the striped block-like hybrid structure is consistently higher than that of the other three steady structures, exhibiting a stronger diffusive tendency. Combined with Fig. 2(a), this further validates the view that the striped block-like hybrid structure is regarded as a transition state between unstable and stable states. In the transverse stripe structure and tic-tac-toe structure, although particles exhibit superdiffusive behavior in the initial stage, the scaling exponent $\alpha$ decreases rapidly over time and even drops below 1, eventually showing subdiffusive behavior. To explain this phenomenon, this study presents the temporal evolution of the scaling exponent $\alpha$ under different ratios $P$ at $D_r = 0.05$, as shown in Fig. 6(a). The results indicate that, in the longitudinal comparison at this stage, as P increases, the curves of $\alpha$ versus time exhibit a trend of first decreasing and then increasing. This trend corresponds to the evolutionary process of the system's structure, which transforms from the transverse stripe structure to the tic-tac-toe structure and finally to the longitudinal stripe structure—meaning the subdiffusive behavior of the tic-tac-toe structure is more prominent. This phenomenon can be attributed to two competing factors. When the ratio $P$ is small (i.e., the longitudinal light intensity is weak), during the transformation of the system's structure from the transverse stripe structure to the tic-tac-toe structure, the proportion of particles trapped in the transverse and longitudinal shadowed regions increases— as shown in Fig. 6(b)— leading to a decrease in the overall diffusive activity of the system and a reduction in the MSD growth rate. However, as $P$ gradually increases, light intensity becomes the dominant influencing factor: although the particle number density in the shadowed regions is relatively high at this point, the particles at the boundary between shadowed regions and illuminated regions acquire high motile activity due to the strong light intensity, which promotes an increase in the particle diffusion rate.

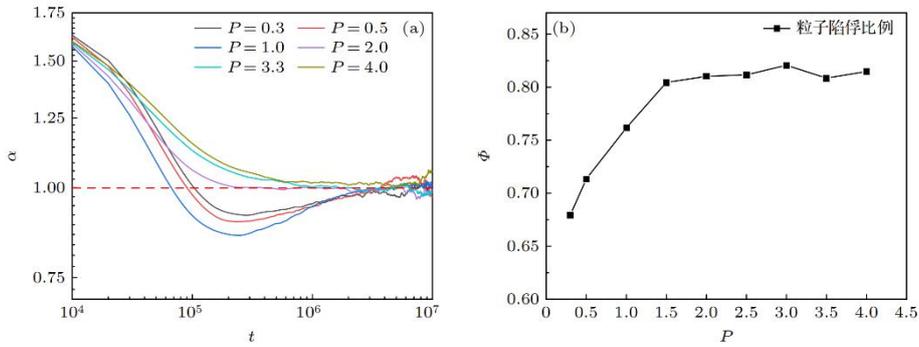

**Figure 6.** (a) Variation of the scaling exponent $\alpha$ with time for different $P$ ratios; (b) particle number density variation in the shadowed region as a function of ratio $P$ at $D_r = 0.05$.

3.3 Effects of Varying Illuminated Region Width ($W$) and Spacing Between Adjacent Illuminated Regions ($S$) on the System's Aggregated Structures

  To further investigate the effect of the light field on the collective behavior of particles, this study systematically varied the width ($W$) of illuminated regions and the spacing ($S$) between adjacent illuminated regions under steady-state conditions; the resulting phase diagram is shown in Fig. 7. As observed in Fig. 7 : when $P = 0.5$, the system primarily exhibits a transverse stripe structure; when $P = 1.0$, the tic-tac-toe structure dominates; and when $P = 1.5$, the system is dominated by a longitudinal stripe structure. This result is consistent with the analytical conclusions from the phase diagram of the system's aggregated structures in Fig. 2(a), further validating the trend of the system's aggregated structures changing with parameters under different light fields. Furthermore, from the overall trend, it can be observed that in the bottom-right corner of the phase diagram—i.e., under the conditions of wider illuminated regions and narrower spacing (corresponding to narrower shadowed regions)—the tic-tac-toe structure is more likely to form. This is because wider illuminated regions provide a larger space for the movement of active particles, while narrower shadowed regions restrict the distribution range of particles, promoting the dense aggregation of particles within these regions. Meanwhile, due to the repulsive interactions between particles, particles cannot fully aggregate in the block-like shadowed regions. It is the combined effect of this geometric confinement and the interactions between high-density particles that causes particles to tend to form the tic-tac-toe structure. Additionally, a new structure— the checkerboard-like structure— appears in the phase diagram; further details are available in the supplementary video SV8 (online). This checkerboard-like structure is more likely to form under conditions of wider spacing, as wider spacing increases the difficulty of particle migration, ultimately leading to particles being trapped in discrete checkerboard grids.

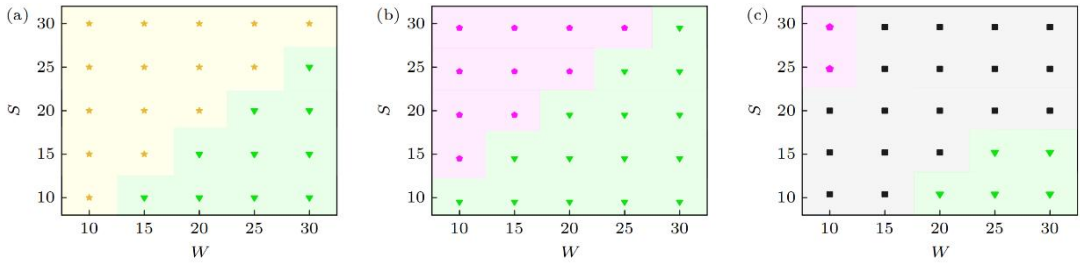

**Figure 7.** Phase diagrams showing the effect of varying the illumination region width $W$ and the spacing $S$ between adjacent illuminated regions on the aggregation structure of the system: (a) $D_r = 0.05$, $P = 0.5$; (b) $D_r = 0.05$, $P = 1.0$; (c) $D_r = 0.05$, $P = 1.5$. ★ represents the transverse stripe structure; ▼ represents the tic-tac-toe structure; ● represents the checkerboard-like structure; ■ represents the longitudinal stripe structure.

## 4. Conclusion

Unlike previous studies that only considered active particles[44,52–56], the particles in the model of this study exhibit different properties in different regions: in illuminated regions with different intensities, they behave as active Brownian particles (ABPs) with varying activities, while in shadowed regions, they behave as ordinary Brownian particles. This study finds that particles aggregate in shadowed regions and weak-illumination regions. This is because the motion of active particles induces phase separation (MIPS): the difference in particle motion velocity under different light intensities causes particles to spontaneously separate into high-density and low-density regions. On this basis, the collective behavior of particle swarms can be regulated by adjusting the rotational diffusion coefficient and the relative intensity of transverse and longitudinal light. When illuminated regions and shadowed regions are uniformly alternating along the x-axis and y-axis, the system exhibits two unstable states and four stable states. The unstable states include the unstable cluster structure and the oscillating structure, while the stable states consist of the transverse stripe structure, the tic-tac-toe structure, the longitudinal stripe structure, and the striped block-like hybrid structure. Among these, the striped block-like hybrid structure can be regarded as a transition state between stable and unstable states. As the rotational diffusion coefficient $D_r$ increases, the particles show a stronger ability to adapt their motion to changes in the light field, and the system as a whole tends to be more stable. With an appropriate rotational diffusion coefficient, gradually increasing the relative intensity of longitudinal light leads to the evolution of the system's structure from the transverse stripe structure to the longitudinal stripe structure. In terms of dynamics, unstable states exhibit sustained superdiffusive behavior throughout the entire simulation process, while stable states show superdiffusion in the initial stage and eventually evolve into normal diffusion. To further investigate the regulatory effect of the light field on particle behavior under stable states, this study systematically varied the width $W$ of illuminated regions and the spacing $S$ between adjacent illuminated regions. It was found that, in addition to the previously identified main structures, the system can also form the checkerboard-like structure under the conditions of wider spacing and narrower illuminated regions. This study focuses on investigating the effects of light intensity, rotational diffusion coefficient, and the width and spacing of illuminated regions on the aggregated structures of the system. It explores the phase separation behavior of particles in a complex light field environment, thereby providing valuable insights for the regulation of aggregated states in active particle systems.